%% file: main_CCTA.tex
\documentclass[letterpaper, 10 pt, conference]{ieeeconf}  % Comment this line out
\IEEEoverridecommandlockouts                              % This command is only
\usepackage{geometry}
\usepackage{amsmath} % assumes amsmath package installed
\usepackage{amssymb}  % assumes amsmath package installed
\usepackage{amsfonts}
\usepackage[subnum]{cases}
\usepackage{multirow}
\usepackage{textcomp}
\usepackage{matlab-prettifier}
\usepackage{listings}

\usepackage{xcolor}
\usepackage{hyperref}
\usepackage{cleveref}
\usepackage{lineno}
\usepackage{graphicx}
\usepackage{float}
\usepackage{svg}
\usepackage{wrapfig}

\title{\LARGE \bf
First experimental demonstration of plasma shape control in a tokamak through Model Predictive Control
}

\author{Adriano Mele$^{1}$, 
Maria A. Topalova$^{1,2}$, 
Cristian Galperti$^{1}$, Stefano Coda$^{1}$, \\
the TCV team$^{3}$ and the Eurofusion Tokamak Exploitation team$^4$% 
\thanks{$^{1}$Ecole Polytechnique Fédérale de Lausanne (EPFL), Swiss Plasma Center (SPC), CH-1015 Lausanne, Switzerland}%
\thanks{$^{2}$ Department of Physics \& Astronomy - School of Natural Sciences, The University of Manchester, Oxford Rd, Manchester, M13 9PL UK}%
\thanks{$^{3}$See author list of B. P. Duval et al 2024 Nucl. Fusion 64 112023}%
\thanks{$^4$See the author list of E. Joffrin et al 2024 Nucl. Fusion 64 112019}
}%
\begin{document}

\maketitle
\thispagestyle{empty}
\pagestyle{empty}

\begin{abstract}
In this work, a Model Predictive Controller (MPC) is proposed to control the plasma shape in the Tokamak à Configuration Variable (TCV). The proposed controller relies on models obtained by coupling linearized plasma response models, derived from the \texttt{fge} code of the Matlab EQuilibrium toolbox (MEQ) suite, with a state-space description of the core TCV magnetic control system. It optimizes the reference signals fed to this inner control loop in order to achieve the desired plasma shape while also enforcing constraints on the plant outputs. To this end, a suitable Quadratic Programming (QP) problem is formulated and solved in real-time. The effectiveness of the proposed controller is illustrated through a combination of simulations and experimental results. To the best of our knowledge, this is the first time that a plasma shape control solution based on MPC has been experimentally tested on a real tokamak.
\end{abstract}

\smallskip

%% main text
\input{sections/introduction}

\input{sections/tcv}
\input{sections/mpc}
\input{sections/results}

\input{sections/conclusions}
\input{sections/acknowledgment}

\bibliographystyle{IEEEtran}
\bibliography{mybib.bib}

\end{document}

%% file: sections/introduction.tex
\section{Introduction}\label{sec:Introduction}
Thermonuclear fusion represents a promising solution for sustainable energy production. However, maintaining a fusion reaction introduces various challenges, one of which is plasma confinement. Potential approaches include inertial and magnetic confinement. In particular, magnetic confinement can be achieved through different devices, such as stellarators and tokamak reactors. The present study focuses on the latter, specifically addressing the problem of magnetic shape control for the TCV tokamak.
Tokamak magnetic control has been recognised as a crucial aspect of ongoing efforts towards viable fusion, leading to its development as a relatively mature field of research \cite{ariola2008, pironti2005}. 
% Within this context, shape control falls under the broader umbrella of axisymmetric magnetic control \cite{tokamak:ACC2020} and constitutes the main objective of this work. 
Previous research \cite{xsc2005,gates2005plasma,ambrosino2020} has established the use of PID controllers combined with decoupling techniques as a standard approach for real-time operations. While effective for basic control tasks, this approach exhibits limitations in handling complex, nonlinear, and multivariable control problems.

The development of the TCV shape control framework dates back to 1998 \cite{ariola1999modern}, when a model-based PID controller coupled with an H-infinity formalism was proposed. Similarly, in~\cite{anand2017} a PI controller is suggested, relying on the feedback signals provided by the RT-LIUQE reconstruction code~\cite{moret2015tokamak} and adopting singular value decomposition (SVD) for decoupling purposes. 
The integration of deep reinforcement learning has also been explored in~\cite{degrave2022}. 
Recent works~\cite{mele2024codit, mele:ssrn} have established an architecture that integrates the existing magnetic reconstruction tools into a shape control scheme based on isoflux and a MIMO PID controller combined with an SVD-based decoupling strategy. The proposed solution is similar to the one in \cite{anand2017}, with the main difference being that it leaves the core magnetic controller fundamentally unchanged.

Moreover, this architecture has been equipped with different current allocation algorithms, similar to those proposed for JET \cite{de2011nonlinear} and DEMO \cite{JOTA_CLA}. These allocation systems aim to solve secondary optimisation problems alongside the main task of controlling the plasma shape. Secondary problems of interest include minimising the quadratic norm of the coil current vector, penalising currents exceeding the saturation limits \cite{CODIT:allocator}, or modifying shape control actions to impose hard constraints, such as coil current saturations \cite{QPalloc}.
These approaches have the advantage of being conceptually decoupled from the existing control architecture.% the specific shape controller implementation might even be unknown and/or inaccessible. 

However, if the controller is accessible, a constrained optimal control solution can provide a comprehensive means of addressing both the primary control objective and various secondary objectives, including hard constraints, while minimising control effort (e.g., coil current demands) in accordance with the appropriate weight embedded in the quadratic problem.
Notably, the concept of Model Predictive Control (MPC) is a popular and well-assessed tool in the existing literature on optimal control
%\cite{rossiter_model-based_2017}. 
\cite{wang_model_2009, rossiter_model-based_2017}. 
Its application to various control problems in Tokamak reactors has been explored, such as core fuelling and density control \cite{orrico2023mixed}, safety factor profile \cite{maljaars2015control}, current density distribution and stored thermal energy \cite{maljaars2015model}, internal inductance and current profiles \cite{garrido2014internal, garrido2016real}. Several attempts at the implementation of MPC algorithms for plasma shape control have also been proposed, relying on different simulation environments.
In \cite{mattei2013constrained}, a constrained, linear, model predictive gap shape controller is proposed in a reference governor scheme for potential application on the ITER Tokamak. Similar designs with additional components, such as disturbance estimation \cite{gerkvsivc2016iter} and input dimensionality reduction through SVD \cite{gerkvsivc2018model}, were developed for the same purpose. Furthermore, the possibility of employing an MPC algorithm for plasma shape control in DEMO has been explored in \cite{tartaglione2022plasma}. 
While all of the above have been extensively validated through simulations, as far as we are aware, the MPC paradigm has never been experimentally tested in real-time Tokamak operation for plasma shape control.
The aim of this paper is to propose a comprehensive solution to the shape control problem in the TCV Tokamak by employing an MPC algorithm and leveraging the existing TCV modelling \cite{carpanese:phd} and control \cite{felici2014simulink, galperti2024overview} framework.

% The rest of this article is organised as follows. Sec.\ref{sec:tcv} provides an overview of the TCV Tokamak and its control system, together with the main modelling assumptions used in this work. Sec.\ref{sec:mpc} describes the design procedure of the proposed MPC shape controller. In Sec.\ref{sec:exp}, the effectiveness of the proposed controller is illustrated by means of both numerical simulations and experimental results. Finally, Sec.\ref{sec:conclusions} concludes the paper.

%% file: sections/tcv.tex
\section{The TCV tokamak}\label{sec:tcv}

% \begin{wrapfigure}{l}
%     \centering
%     \includegraphics[width=0.4\linewidth, trim=10 10 10 0]{img/tcv_layout.eps}
%     \caption{TCV poloidal cross-section, with a sample equilibrium from pulse \#78071. The OH, E, F, and G coils are shown in yellow, green, blue and red, respectively. Red crosses show the flux loops position, while black dashes indicate the pick-up coils. The green markers show a possible choice of shape control points (figure adapted from~\cite{mele2024codit}).}
%     \label{fig:tcv_layout}
% \end{wrapfigure}

\begin{figure}[h]
    \centering
    \includegraphics[width=0.4\linewidth, trim=10 10 10 0]{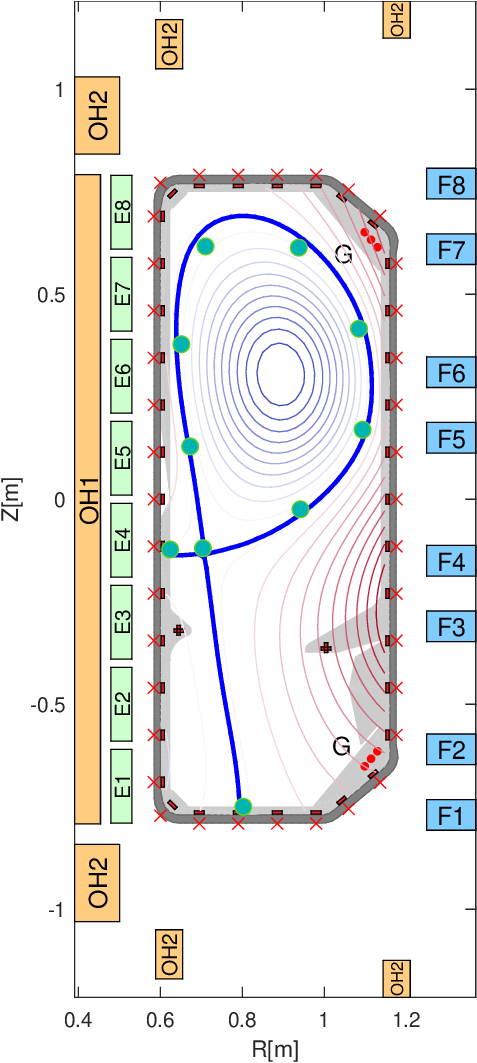}
    \caption{TCV poloidal cross-section, with a sample equilibrium from pulse \#78071. The OH, E, F, and G coils are shown in yellow, green, blue and red, respectively. Red crosses show the flux loops position, while black dashes indicate the pick-up coils. The green markers show a possible choice of shape control points (figure adapted from~\cite{mele2024codit}).}
    \label{fig:tcv_layout}
\end{figure}

TCV~\cite{hofmann1994, duval2024experimental} is a medium-size tokamak ($R_0 = 0.88$m) operated by the Swiss Plasma Center (SPC) of the École Polytechnique Fédérale de Lausanne. 
It is equipped with a set of $19$ independent Poloidal Field (PF) circuits, conceptually divided into four sets: the $OH1-2$ coils, dedicated to the control of the plasma current $I_p$, the $E1-8$ and $F1-8$ coils, used to control plasma position and shape, and the $G$ coils, located inside the vacuum vessel, which are employed by the vertical stabilisation controller. The layout of the coils is shown in Fig.~\ref{fig:tcv_layout}. The high number of independent circuits installed in TCV provides it with a unique flexibility in terms of plasma shaping, making the investigation of different magnetic geometries a core research mission for this device.
 
The core magnetic controller of TCV is a MIMO, PID-based linear controller called \emph{hybrid}, which takes care of vertically stabilising the plasma and controlling its current and position. It also features a density control channel, neglected in this work for simplicity. More details on the hybrid system can be found in~\cite{lister1998high}.
A fully digital version of this controller is available in the digital control system of TCV, also called the Système de Contrôle Distribué (SCD)\cite{le2014,galperti2024overview}. 
% In particular, since hybrid is a fully linear controller, a state-space representation can be obtained for it. In the following, it is assumed that such representation is available and given by
In the following, it is assumed that a state-space representation is available for the hybrid controller in the form
\begin{equation}
\begin{aligned}
    \dot{x}_h(t) &= A_{h} x_h(t) + B_{h} e_h(t) \\
          V_a(t) &= C_{h} x_h(t) + D_{h} e_h(t)\,,
\end{aligned}  \label{eq:hybss}
\end{equation}
where $x_h \in \mathbb{R}^{n_h}$ is the controller state, $e_h=r_h-y_h \in \mathbb{R}^{m_h}$ is the vector of error signals input to the hybrid controller, and $V_a \in \mathbb{R}^{19}$ are the voltage requests to the active coils generated by the controller. The quantities controlled by hybrid include two channels devoted to the control of the plasma current centroid position coordinates $rI_p$ and $zI_p$, and $14$ channels devoted to the control of the $I_{EF}$ currents projected onto a subspace which is orthogonal with respect to the directions used for position control, denoted $I_{orth}$ in what follows (see~\cite{mele:ssrn} for details).
The hybrid controller has recently been equipped with an external shape control loop, which acts on its reference signals.
% Closed-loop shape control relies on the availability of a real-time equilibrium reconstruction, provided in TCV by the LIUQE code~\cite{moret2015tokamak}. 
As discussed in~\cite{mele2024codit,mele:ssrn}, the shape controller of TCV is based on the \emph{isoflux} approach~\cite{ferron1998real}, and controls the magnetic field or the poloidal flux at a set of user-defined control locations in order to force the Last Closed Flux Surface, which defines the shape of the plasma, to pass through such points. As an example, the single-null plasma configuration used for the experiments discussed in this article is shown in Fig.~\ref{fig:tcv_layout}. For this configuration, the control objective is to regulate to zero both the poloidal flux differences between the selected boundary points and the X point and the magnetic field components at the X point. The necessary feedback signals are obtained by interpolating the real-time equilibrium maps generated by the LIUQE code~\cite{moret2015tokamak} at the specified locations. 
% The controlled quantities are defined as linear combination of the poloidal flux and magnetic field values relative to the zero-flux reference, known as the X-point. They are obtained by interpolating the real-time equilibrium maps generated by the LIUQE code~\cite{moret2015tokamak}. 

%The specific quantities that are controlled depend on the considered plasma configuration; for example, in a single-null plasma such as the one in Fig.~\ref{fig:tcv_layout}, the magnetic field at the desired X-point and the difference in poloidal flux between a set of target boundary points and the target X-point are all controlled to zero. 

%A way to interpret isoflux feedbacks in terms of plasma shape displacements is discussed in~\cite{tenaglia2024interpretable}. 

\begin{figure*}[t]
   \centering
   \includegraphics[width=0.85\linewidth, trim = 0 10 0 0]{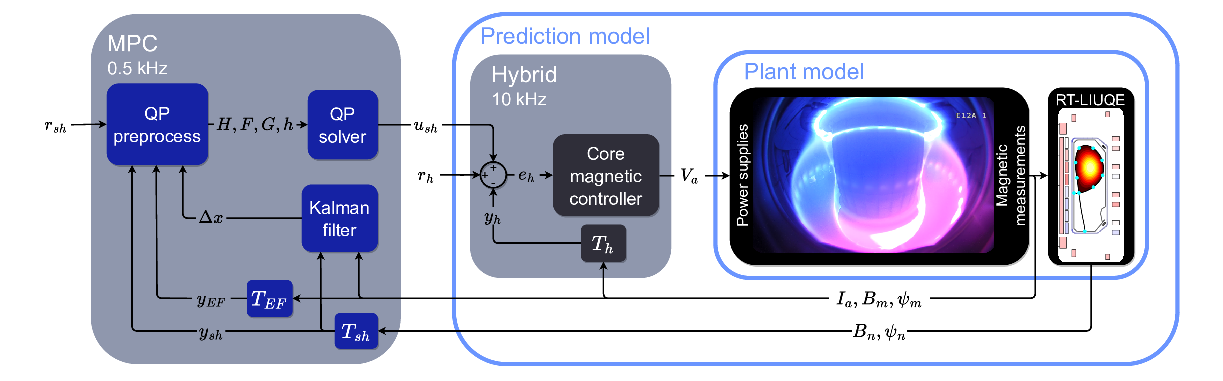}
   \caption{Simplified view of TCV magnetic controller. The core hybrid block, running at $10$~kHz, vertically stabilises the plasma and controls its current and position, along with the currents in the PF coils. The outer loop, devoted to shape control and running at a lower sampling frequency, takes virtual flux and fields measurements produced by the real-time LIUQE reconstruction algorithm, transformed into the desired controlled quantities through the $T_{sh}$ block. The output of such controller are modifications to the references to hybrid. The figure also shows the blocks included in the plant model (generated by \texttt{fge}) and in the MPC prediction model.}
   \label{fig:tcv-ctrl}
\end{figure*}

The design of plasma shape controllers is often based on a reference plasma equilibrium and relies on the availability of linearized models of the dynamics of the plant in the vicinity of the target configuration. The axisymmetric magnetohydrodynamic (MHD) equilibrium of a tokamak plasma can be described by an elliptic PDE known as the Grad-Shafranov equation, which can be linearized and coupled with a model of the evolution of the currents in the coils and passive conductive structures surrounding the plasma~\cite{ariola2008}.
In particular, a linearized model of the TCV plasma response around a desired target configuration was obtained in this work through the linearization module of the \texttt{fge} code~\cite{carpanese:phd}.
The resulting linearized model can be written as
\begin{equation}
\begin{aligned}
    \delta\dot{x}_{p}(t) &= A_{p} \delta x_{p}(t) + B_{p} V_a(t) + E_p \delta \dot{w}(t) \\
    \delta y_{p}(t) &= C_{p} \delta x_{p}(t) + F_p \delta w(t)\,,
\end{aligned}  \label{eq:plasmass}
\end{equation}

where $\delta x_p \in \mathbb{R}^{n_p}$ is the plant state, consisting of the currents flowing in the PF coils, in the passive elements and in the plasma, while the vector $\delta y_p$ contains the outputs of interest, including the magnetic field and flux $B_m, \psi_m$ measured by the available probes, the active currents $I_a$, and the magnetic flux and field at the shape control points $B_n, \psi_n$. All quantities denoted by $\delta$ in~\eqref{eq:plasmass} are intended as deviations from the equilibrium reference values. 
The vector~$w$ contains variations of parameters describing the internal plasma current distribution; often the poloidal beta and the internal inductance are used, i.e.~$w = [\beta_p, l_i]^T$, as discussed in~\cite{ariola2008}. 
For reference, the elements included in the plant model are represented in a blue box in Fig.~\ref{fig:tcv-ctrl}. 

With reference to the plant model~\eqref{eq:plasmass}, let the hybrid-controlled quantities $y_h \in \mathbb{R}^{m_h}$ be related to the output vector $y_p$ through the linear transformation $y_h = T_{h} y_p$, and the quantities $y_{sh} \in \mathbb{R}^{l_{sh}}$ controlled by the shape controller through $y_{sh} = T_{sh} y_p$. The matrix $T_{h}$ contains the linear observers used to obtain feedback signals for the hybrid controller (see discussion in~\cite{mele:ssrn}), while typically $T_{sh}$ is used to select the field components for the controlled X-point(s) and to subtract the flux at the target primary X-point from the flux at the other shape control points. Moreover, let us explicitly denote by $y_{EF} = T_{EF} y_p$ the outputs associated to the currents in the $E$ and $F$ coil sets (see Fig.~\ref{fig:tcv_layout}), i.e. the ones used by the shape controller on TCV. Note that, $y_{EF} \subset I_a$ with $y_{EF} \in \mathbb{R}^{l_{EF}}$, where $l_{EF}=16$.

For simplicity, in the preliminary design discussed in this work it is assumed that the plasma internal profiles are fixed during the MPC control window, i.e. $\delta w = \delta \dot{w} = 0$. Under this assumption, a linearized model of the whole plant as seen by the shape controller can be obtained by interconnecting the two state-space models above as
{\small
\begin{equation} \label{eq:mpcss}
\begin{aligned}
    \delta\dot{x}_p(t) &= A_{p} \delta x_p(t) + B_{p} (C_{h} x_h(t) + D_{h} (r_h - T_h C_{p} \delta x_{p}(t)) ) \\    
    \dot{x}_h(t) &= A_{h} x_h(t) + B_{h} (r_h - T_h C_{p} \delta x_{p}(t)) \\
    V_a(t) &= C_{h} x_h(t) + D_{h} (r_h - T_h C_{p} \delta x_{p}(t)) \\
    \delta y_{p}(t) &= C_{p} \delta x_p(t) \,,
\end{aligned} 
\end{equation}
}
It is worth noticing that the equation describing the evolution of $V_a$ in~\eqref{eq:mpcss} can be used to easily integrate voltage constraints in the proposed MPC scheme. This is not investigated in the present work and is left for future developments.

%% file: sections/mpc.tex
\section{MPC design}\label{sec:mpc}
The proposed MPC scheme aims to control the shape descriptors $y_{sh}$ to a desired reference while simultaneously imposing constraints on the other outputs of interest. For the design proposed in this work, saturations are enforced on the $I_{EF}$ currents. The design rationale behind the controller was to operate as an outer loop over the inner hybrid control loop, as shown in Fig.~\ref{fig:tcv-ctrl}, and directed towards overcoming the challenges of real-time implementation.

\subsection{Prediction model}

% The proposed MPC scheme aims to control the shape descriptors $y_{sh}$ to a desired reference while also imposing constraints on other outputs of interest. In particular, in the proposed examples, saturations are enforced on the $I_{EF}$ currents.
% 
To derive a prediction model for the controller, let us denote by $x = [x_p^T, x_h^T]^T$ the state of the closed-loop system~\eqref{eq:mpcss} and by $y=[T_{EF}^T, T_{sh}^T]^T y_{p}$ the vector containing the outputs of interest. 
The following state-space matrices shall be considered
\begin{equation}
\begin{aligned}
    A &= 
    \begin{bmatrix}
        A_p - D_h T_h C_p &B_p C_h\\
        -B_h T_h C_p      &A_h
    \end{bmatrix}\,, \quad
    &B &= 
    \begin{bmatrix}
         D_h \\
         B_h
    \end{bmatrix}\,, \\
    C &= 
    \begin{bmatrix}
         T_{EF} C_p & \mathbf{0}_{[l_{EF} \times n]} \\
         T_{sh} C_p & \mathbf{0}_{[l_{sh} \times n]} 
         
    \end{bmatrix} \,, \quad
    &D &= \mathbf{0}_{[l \times n]} \,.
\end{aligned}\label{eq:mpcmatrices}
\end{equation}
{Here and in the following, $\mathbf{0}_{[a,b]}$ denotes a zero matrix of dimension $a \times b$ and $\mathbf{I}_c$ denotes the identity matrix of dimension $c \times c$, with $a,b,c \in \mathbb{N}$. In~\eqref{eq:mpcmatrices}, $n = n_p + n_h$ and $l = l_{EF} + l_{sh}$. For brevity, the indication of the dimension is dropped when clear from the context.}

The system described by the state-space matrices in~\eqref{eq:mpcmatrices} is then discretised, assuming a zero-order hold between consecutive samples.
% The matrices of the discrete-time, reduced model will be denoted by $A,\,B,\,C,\,D$. 
In what follows, the state, output and input vectors of the prediction model at time $t_k$ will be denoted by $x_k \in \mathbb{R}^{n}$, $y_k \in \mathbb{R}^{l}$, $u_k \in \mathbb{R}^{m}$, respectively, while the references for the shape control at the same time by $r \in \mathbb{R}_l$. In the proposed MPC implementation, $m = n_{EF} < m_h$, i.e. the shape controller acts on a subset of the hybrid references, in particular on those with a direct connection to the $E$ and $F$ coil currents. 
Moreover, the MPC prediction and control horizon will be referred to as $N$ and $N_{c}$, respectively; for a generic signal $z(t) : \mathbb{R} \to \mathbb{R}^p,\, p \in \mathbb{N}$, we denote the sequence of samples from time $t_k$ to time $t_{k+N}$ in vectorized form as
% we denote by z[k,k+T], where
% k ∈ Z, T ∈ N, the restriction in vectorized form of z to the
% interval [k, k + T] ∩ Z, namely
\begin{equation}\label{eq:MPCpred}
    z_ {[k,k+N]} =
    \begin{pmatrix}
        z_{k}   \\
        z_{k+1} \\
        z_{k+2} \\
        \vdots  \\
        z_{k+N}
    \end{pmatrix} \,.
\end{equation}
To achieve offset-free tracking {\cite{rossiter_model-based_2017}}, the linear model~\eqref{eq:mpcss} is recast in velocity form by setting
$ \Delta x_{k} = x_k - x_{k-1}$, $\Delta u_{k} = u_k - u_{k-1} $.
% to obtain
% \begin{equation}
%     \begin{aligned}
%         \Delta x_{k+1} &= A \Delta x_{k} + B \Delta u_{k}\\
%        y_{k} &= C{ \Delta x_{k} } + D \Delta u_{ k } + y_{k-1} \,.
%     \end{aligned}
% \end{equation}
% 
By defining the following matrices
\footnote{Note that, in the general formulation of the prediction model~\eqref{eq:predmat}, the $D$ matrix has been kept. While it is zero in the considered case, it may be retained if voltage saturations are considered ($D_h \neq 0$) or profile modifications are taken into account ($F\neq 0$, with a suitable re-definition of the input vector).}
\begin{equation}\label{eq:predmat}
\begin{aligned}
    &\Psi = 
        \begin{bmatrix}
        C         \\
        C+CA      \\
        %C+CA+CA^2 \\
        \vdots    \\
        \sum\limits_{i=0}^{N}CA^{i} 
        \end{bmatrix}\,, \qquad    
    \Omega_l
    = 
        \begin{bmatrix}
        \mathbf{I}_l \\
        \mathbf{I}_l \\
        \vdots        \\
        \mathbf{I}_l
        \end{bmatrix} \,,  \\
    &\Phi = 
     {\footnotesize % scriptsize
        \begin{bmatrix}
        D                           & \mathbf{0}                  & \dots  & \mathbf{0} \\
        D+CB                        & D                           & \dots  & \mathbf{0} \\
        D+CB+CAB                    & D+CB                        & \dots  & \mathbf{0} \\
        \vdots                      & \vdots                      & \ddots & \vdots     \\
        D + \sum\limits_{i=0}^{N-N_c}CA^{i}B & D + \sum\limits_{i=0}^{N-N_c-1}CA^{i}B & \dots  & D
        % D+CB+\dots+CA^{N-1}B & D+CB+\dots+CA^{N-2}B & \dots & \dots & D
        \end{bmatrix}\,,
    } 
\end{aligned}
\end{equation}
% 
% \begin{equation}\label{eq:Phi}
%     \Phi = 
%     % {\tiny % scriptsize
%     %     \begin{bmatrix}
%     %     D                           & \mathbf{0}                  & \mathbf{0} & \dots  & \mathbf{0} \\
%     %     D+CB                        & D                           & \mathbf{0} & \dots  & \mathbf{0} \\
%     %     D+CB+CAB                    & D+CB                        & D          & \dots  & \mathbf{0} \\
%     %     \vdots                      & \vdots                      & \vdots     & \ddots & \vdots     \\
%     %     D + \sum_{i=0}^{N-1}CA^{i}B & D + \sum_{i=0}^{N-2}CA^{i}B & \dots      & \dots  & D
%     %     % D+CB+\dots+CA^{N-1}B & D+CB+\dots+CA^{N-2}B & \dots & \dots & D
%     %     \end{bmatrix}\,,
%     % }
%      {\footnotesize % scriptsize
%         \begin{bmatrix}
%         D                           & \mathbf{0}                  & \dots  & \mathbf{0} \\
%         D+CB                        & D                           & \dots  & \mathbf{0} \\
%         D+CB+CAB                    & D+CB                        & \dots  & \mathbf{0} \\
%         \vdots                      & \vdots                      & \ddots & \vdots     \\
%         D + \sum_{i=0}^{N-1}CA^{i}B & D + \sum_{i=0}^{N-2}CA^{i}B & \dots  & D
%         % D+CB+\dots+CA^{N-1}B & D+CB+\dots+CA^{N-2}B & \dots & \dots & D
%         \end{bmatrix}\,,
%     }
% \end{equation}
 
% \begin{equation}\label{eq:omega}
%     \Omega
%     = 
%     \begin{bmatrix}
%     \mathbf{I}_l, 
%     \mathbf{I}_l, 
%     \dots, 
%     \mathbf{I}_l
%     \end{bmatrix}^T \,.
% \end{equation}
% 
\noindent
a prediction model for the MPC controller can be written as
\begin{equation}\label{eq:MPCpred}
\begin{aligned}
    y_{[k,k+N]}
    &= \Psi \Delta x_k + \Phi \Delta u_{[k,k+N_{c}]} + \Omega_l y_{k-1} \\
    &= \hat{y}_{[k,k+N]} + \Phi \Delta u_{[k,k+N_{c}]} \,,
\end{aligned}
\end{equation}
where $\hat{y}_{[k,k+N]}$ denotes the predicted output when no control variations are given over the prediction horizon with respect to $u_{k-1}$. 

%Note that, to obtain the predicted outputs $y_{[k,k+N]}$, the term $\Delta x_k$ is needed; to compute this term, the state at time $t_k$ and $t_{k-1}$ must be known. For this reason, the proposed controller has been equipped with a Kalman filter designed on the basis of a reduced version of the closed-loop model~\eqref{eq:mpcss}. 
In the term $\Delta x_k$, the state variables $x_k$ and $x_{k-1}$ are estimated by means of a Kalman filter, based on the closed-loop model~\eqref{eq:mpcss}. The filter receives the measured plant outputs $y_p$ and the hybrid references modified by the shape control actions $r_h + u_{sh}$. Since model~\eqref{eq:mpcss} is a linearization in the vicinity of a reference equilibrium, the output deviations $\delta y_p$ are obtained by subtracting from $y_p$ the output values sampled at the switch-on time of the MPC controller. Upon activation of the MPC, the Kalman filter is run for a few steps before the execution of the real-time optimizer starts.
In practice, both the prediction model~\eqref{eq:predmat} and the Kalman filter are based on a reduced version of~\eqref{eq:mpcss}, obtained through a Hankel balanced truncation with a reduced state dimension of $n=30$.

\subsection{Optimization problem formulation}\label{sec:MPCopt}

At each time step~$t_k$, the proposed MPC solves a QP optimization problem in the standard form
\begin{equation}\label{eq:MPCopt}
\begin{aligned}
    % \min_{\Delta u_{[k,k+N]}} \qquad &J(r_{[k,k+N]}, \Delta u_{[k,k+N]}) \\
    \min_{\Delta u_{[k,k+N_{c}]}} \qquad & \Delta u_{[k,k+N_{c}]}^T H \Delta u_{[k,k+N_{c}]} + F^T \Delta u_{[k,k+N_{c}]} \\
    \text{s.t.} \qquad
    &G {\Delta u_{[k,k+N_{c}]}} \le h \,.
\end{aligned}
\end{equation}
To obtain an optimization problem in this form, consider the cost function
{\small
\begin{equation}\label{eq:MPCcost}
\begin{aligned}
    J&(r_{[k,k+N]}, \Delta u_{[k,k+N_{c}]}) = \\
    &(r_{[k,k+N]}-y_{[k,k+N]})^T W (r_{[k,k+N]}-y_{[k,k+N]}) \\
    &+ \Delta u_{[k,k+N_{c}]}^T Q \Delta u_{[k,k+N_{c}]} = \\
    % &= 
    % (r_{[k,k+N]} - \Psi \Delta x_k - \Phi \Delta u_{[k,k+N]} - \Omega y_{k-1})^T W (r_{[k,k+N]} - \Psi \Delta x_k - \Phi \Delta u_{[k,k+N]} - \Omega y_{k-1})
    % +
    % \Delta u_{[k,k+N]}^T Q \Delta u_{[k,k+N]} \\    
    &(\hat{e}_{[k,k+N]} - \Phi \Delta u_{[k,k+N_{c}]})^T W (\hat{e}_{[k,k+N]} - \Phi \Delta u_{[k,k+N_{c}]}) \\
    &+\Delta u_{[k,k+N_c]}^T Q \Delta u_{[k,k+N_{c}]} \,,    
\end{aligned}
\end{equation}
}
where $\hat{e}_{[k,k+N]} = r_{[k,k+N]} - \hat{y}_{[k,k+N]}$, $Q$ and $W$ denote the weights assigned on the control actions and errors, respectively.
We can define the Hessian matrix and the linear term coefficient in~\eqref{eq:MPCopt} as
\begin{equation}\label{eq:hessian}
    H = \Phi^T W \Phi + R \,, \quad F = -W \hat{e}_{[k,k+N]} \,.
\end{equation}
Terms in $J$ that do not depend on the optimization variables~$\Delta u_{[k,k+N]}$ are dropped.
Note that, once the prediction model is fixed, the Hessian can be precomputed, while the $F$ term needs to be updated at each optimization step. 
Specific channels, usually the ones in $y_{sh}$, can be selected by choosing non-zero weights only in the corresponding entries of the weight matrix $W$. Alternatively, small weights can be assigned to the $y_{EF}$ channels along with zero references; this way, the controller will try to find a trade-off between the accuracy of the plasma shape and the square norm of the $I_{EF}$ current vector.

The inclusion of output constraints in the considered MPC design is straightforward. In the examples proposed in this work, we consider the case where saturation limits are imposed on the currents in the $E$ and $F$ coils
\begin{equation}\label{eq:constr}
     \underline{I}_{EF}  \le y_{EF_j} \le \bar{I}_{EF}, \quad j \in \{k, k+1, ... ,k+N\} \,.
\end{equation}
With reference to~\eqref{eq:mpcmatrices}, where the $EF$ coil currents are the first $n_{EF}$ entries in the output vector, we can define the matrices
\begin{equation}
\begin{aligned}
&\gamma_{EF} = 
    \begin{bmatrix} 
         \mathbf{I}_{n_{EF}} &\mathbf{0}_{[n_{EF},\, l-n_{EF}]}
    \end{bmatrix} \,, \\
&\Gamma =   
    \begin{bmatrix} 
        \gamma_{EF}  &\mathbf{0}   & \dots  &\mathbf{0} \\
        \mathbf{0}   &\gamma_{EF}  & \dots  &\mathbf{0} \\
        \vdots       &\vdots       & \ddots &\vdots     \\
        \mathbf{0}   &\mathbf{0}   & \dots  &\gamma_{EF} \,.
    \end{bmatrix} \,, 
\end{aligned}
\end{equation}
The predicted $EF$ currents are given by
\[
y_{EF_{[k,k+N]}} = \Gamma (\Psi \Delta x_k  +  \Phi \Delta u_{[k,k+N_{c}]}  +  \Omega_l y_{k-1}) \,.
\]
Applying the constraint~\eqref{eq:constr} over the prediction horizon, the terms $G$ and $h$ in~\eqref{eq:MPCcost} can be defined as
\begin{equation}
G = 
    \begin{bmatrix}
        +\Gamma \Phi \\
        -\Gamma \Phi
    \end{bmatrix} \,, \quad
h = 
    \begin{bmatrix}
        -\Gamma \hat{y}_{[k,k+N]} + \Omega_{n_{EF}}      \bar{I}_{EF} \\
        +\Gamma \hat{y}_{[k,k+N]} - \Omega_{n_{EF}}\underline{I}_{EF}
    \end{bmatrix} \,,
\end{equation}
where $\Omega_{n_{EF}}$ is defined similarly to $\Omega_l$ in~\eqref{eq:predmat}.

% 
% Applying the constraint~\eqref{eq:constr} over the prediction horizon we obtain
% % 
% \begin{equation*}
% \begin{aligned}
%       \Gamma \Phi \Delta u_{[k,k+N]} &\le - \Gamma \hat{y}_{[k,k+N]} + \Omega_{n_{EF}}\bar{I}_{EF}  \\
%     - \Gamma \Phi \Delta u_{[k,k+N]} &\le + \Gamma \hat{y}_{[k,k+N]} - \Omega_{n_{EF}}\underline{I}_{EF} \,,
% \end{aligned}
% \end{equation*}
% where $\Omega_{n_{EF}}$ is defined similar to $\Omega_l$ in~\eqref{eq:predmat}.
% % 
% Finally, we can define the terms $G$ and $h$ in~\eqref{eq:MPCcost} as
% \begin{equation}
% G = 
%     \begin{bmatrix}
%         +\Gamma \Phi \\
%         -\Gamma \Phi
%     \end{bmatrix} \,, \quad
% h = 
%     \begin{bmatrix}
%         -\Gamma \hat{y}_{[k,k+N]} + \Omega_{n_{EF}}      \bar{I}_{EF} \\
%         +\Gamma \hat{y}_{[k,k+N]} - \Omega_{n_{EF}}\underline{I}_{EF}
%     \end{bmatrix} \,.
% \end{equation}

% \begin{equation}
%     \begin{aligned}
%     \min_{\Delta u_{[k,k+N]}} \qquad &\Delta u_{[k,k+N]}^T H \Delta u_{[k,k+N]} + F^T \Delta u_{[k,k+N]} \\
%     \text{s.t.} \qquad
%     &A {\Delta u_{[k,k+N]}} \le b \,.
% \end{aligned}
% \end{equation}

% \textcolor{red}{Before concluding this section, it is worth to remark that~\eqref{eq:constr} only provide one possible example of output constraints. Another common choice is to constrain the voltages applied to the coils; this can be done in the proposed framework by resorting to the expression for $V_a$~\eqref{eq:mpcss}.}

% in  the expression of the voltages applied to the active coils in~\eqref{eq:mpcss} can be used to impose constraints on the control voltages as well.}

%% file: sections/results.tex
\section{Results and Discussion }\label{sec:exp}
To validate the proposed algorithm, both nonlinear simulations using the \texttt{fge} code~\cite{carpanese:phd} and preliminary experiments on the TCV tokamak were performed on a diverted single-null plasma configuration, i.e., a topology with one X point whose open field lines terminate at strike points on dedicated divertor plates, carrying heat and particles away from the vessel wall (e.g. Fig.~\ref{fig:tcv_layout}). The controller was active during the flat-top phase (from $0.6$ to $1.2$ s), viz., the steady phase of a discharge that follows current ramp-up, when plasma current, shape and density are held roughly constant. Ten boundary control points were chosen, along with two strike points and one primary X-point. 
%A sampling time of $T_s = 2$~ms was selected, assuming a zero-order hold between consecutive samples; this sampling time proved sufficiently long for real-time execution of the proposed algorithm. 
%The Kalman filter was run for $10$~control steps before enabling the RT-optimiser at $0.62$~s. 
A modified version of the primal-dual QP algorithm implementation proposed in~\cite{quadprogPP}, based on~\cite{goldfarb1983numerically}, was used in the experiments. 
A prediction and control horizon of $N=15$ and $N_c=3$ were chosen, leading to an optimisation problem with $48$ decision variables and $480$ constraints to be solved at each time step. A maximum number of $15$ iterations was imposed; benchmarking the real-time solver on randomly generated QPs of the target size yielded a worst-case solve time of $0.48$ ms, well within the chosen $2$ ms sampling period $T_s$, leaving ample margin for I/O and Kalman filtering while keeping the discrete-time phase lag negligible at the target bandwidth. The Kalman filter was run for $10$~control steps before enabling the RT-optimiser at $0.62$~s.

%. This was found to be compatible with the sampling time of $T_s = 2$~ms, chosen by benchmarking the solver employed in the real-time control system on randomly generated problems, which resulted in $0.48$ ms for the allowed iterations maximum and fixed problem size.
% To keep the execution time of the algorithm below $2$~ms, a prediction horizon of $N=15$ was chosen, while the control horizon of the algorithm was chosen as $N_c=3$. Given this selection, a problem with $48$ optimisation variables and $480$ constraints has to be solved at each time step. A maximum number of $15$ iterations was fixed, which was found to be compatible with the chosen sampling time by testing the real-time implementation of the algorithm on randomly generated problems.

\begin{figure}[h] 
    \centering
    \includegraphics[width = 0.75\linewidth, trim= 0 20 0 10]{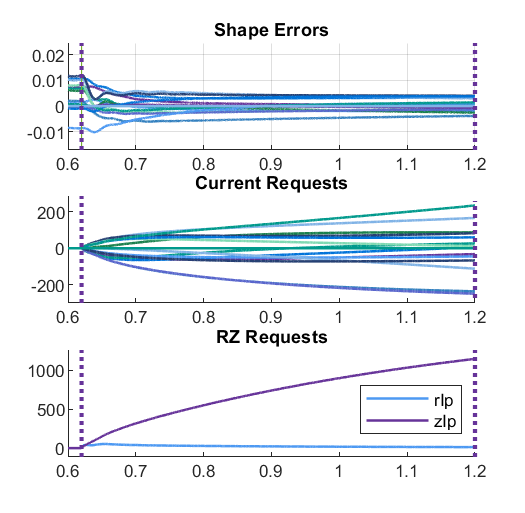}
    \caption{{Simulation of a discharge based on TCV pulse \#84513.}}
    \label{fig:sim-standard}
\end{figure} 

\begin{figure}[h] 
    \centering
    \includegraphics[width = 0.75\linewidth, trim= 0 20 0 10]{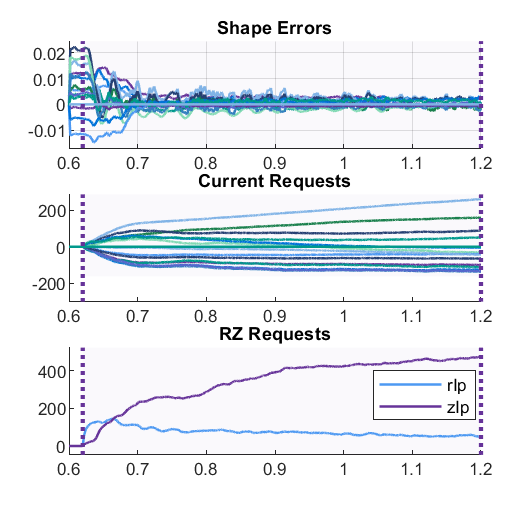}
    \caption{TCV pulse \#84535.}
    \label{fig:exp-1}
\end{figure}

In Figs.~\ref{fig:sim-standard} and~\ref{fig:exp-1} the considered plasma scenario is compared in simulation and a dedicated experiment. The controller was tuned with a weight of $10^7$ on the shape errors, $0$ on the $EF$ coil currents, $30$ and $3$ on the requested variations of $I_{orth}$ and both the $rI_p$ and $zI_p$, respectively. 
The first subplots show the resulting shape control errors, followed by the control requests in terms of the orthogonal current directions $I_{orth}$, as well as $rI_p$ and $zI_p$. A settling time of less than $0.1$~s is observed in both the simulated and experimental results. 
Despite their overall consistency, some discrepancies can be noted. A slightly higher initial error is observed in the experiments, along with residual oscillations in the controlled variables. The latter may be attributed to an undesired interaction with the vertical stabilisation controller, which is not present in simulations and might be ameliorated by retaining a higher number of state variables in the model reduction procedure. The difference in the current requests, more specifically around $0.75$~s, can be explained with a non-ideality of the $OH$ coils power supplies when the currents cross the zero value. It was observed that this event leads to a small plasma displacement towards the desired shape, which results in smaller control requests compared to the simulated case.

\begin{figure}[h]
    \centering
    \includegraphics[width = 0.75\linewidth, trim= 0 20 0 10]{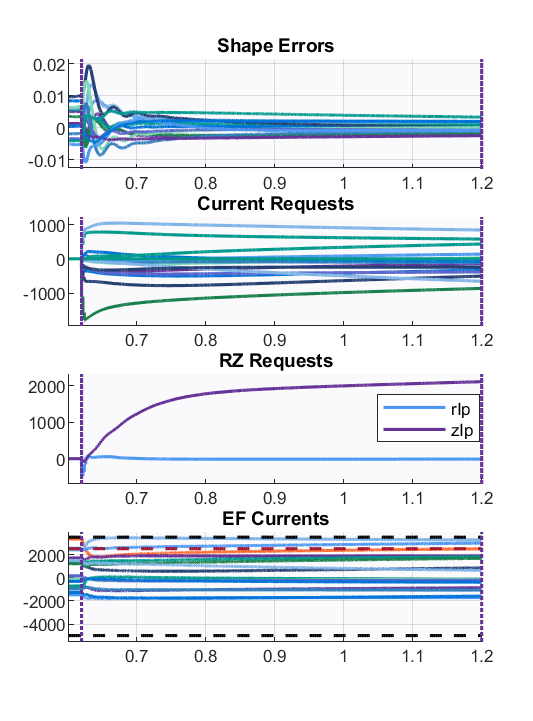}
    \caption{Simulation of a discharge based on TCV pulse \#84513. An artificial saturation of $2500$~A is introduced on the current in coil $E5$, shown in red.}
    \label{fig:sat-sim}
\end{figure}

Subsequently, a simulated case of current saturation is shown in Fig.~\ref{fig:sat-sim}. All parameters remain unchanged, except for the weight on the shape, which was increased by one order of magnitude. Despite the current saturation, the projected errors, as well as the settling time, are comparable to those in the case discussed previously. Additionally, the total $EF$ coil currents are shown in the fourth subplot, where the black dashed line represents a $4$~kA saturation imposed on all currents. An exception is the $E5$ coil current, depicted in orange, for which an artificial saturation limit of $2.5$~kA was chosen, as shown by the red dashed line. These results demonstrate the controller’s ability to manage current saturation effectively while redistributing the remaining currents and adhering to the overall constraints. 

%% file: sections/conclusions.tex
\section{Conclusion}\label{sec:conclusions}

In this article, the first experimental demonstration of MPC for plasma shape control, specifically on the TCV Tokamak, has been discussed. The proposed controller relies on a prediction model that encompasses both the core TCV magnetic control system and a linearised plasma response model, and successfully regulates plasma shape while enforcing physical constraints on selected currents. Experimental results validate the feasibility of the real-time implementation, showing strong agreement with simulations despite minor performance discrepancies. A more extensive validation of the proposed control algorithm on different plasma configurations and in different control scenarios (such as fast plasma displacements or multiple simultaneous current saturations) is expected in the near future, depending on the availability of experimental time on TCV.

The controller can be refined and extended in several directions. For instance, in the current implementation, variations in the internal profiles and in the total plasma current are neglected. These can be easily included in the prediction model in terms of synthetic parameters such as $\beta_p, l_i$, as is common practice in plasma magnetic control~\cite{tartaglione2022plasma}. Moreover, the introduction of SVD modes could be explored to reduce the input dimension and increase the control horizon. The prediction model can be refined to account for residual unmodelled dynamics. Additionally, strategies to update the model and the optimisation problem in real-time could be investigated, possibly resorting to surrogate models of the plasma dynamics derived through Machine Learning approaches.

% {\color{red}
% \begin{itemize}
%     %\item Validate the proposed architecture through simulation in different cases, e.g. with saturated coils, shape variations, different plasma configurations, etc.
%     %\item discuss preliminary experiments performed at the end of 2024
%     %\item remember to add a note on the weights used in the optimization
%     \item one thing to note is that, in the current implementation, variations in the internal profiles and in the total plasma current are neglected. These can be easily included in the prediction model in terms of synthetic parameters such as $\beta_p, l_i$, as it is common practice in plasma magnetic control~\cite{tartaglione2022plasma}. We could mention that the focus of this work is on a first, working implementation, but all of this refinements are foreseen in the (near) future.
% \end{itemize}
% }

%% file: sections/acknowledgment.tex
\section*{Acknowledgments}\label{sec:acks}
{\footnotesize
This work has been carried out within the framework of the EUROfusion Consortium, partially funded by the European Union via the Euratom Research and Training Programme (Grant Agreement No 101052200 — EUROfusion), and The Nuclear Technology Education Consortium (NTEC). The Swiss contribution to this work has been funded by the Swiss State Secretariat for Education, Research and Innovation (SERI). Views and opinions expressed are however those of the author(s) only and do not necessarily reflect those of the European Union, the European Commission or SERI. Neither the European Union nor the European Commission nor SERI can be held responsible for them. This work was supported in part by the Swiss National Science Foundation.
}